\documentstyle[12pt,epsf]{article}

\def\DRAFT{0}
\renewcommand{\theequation}{\thesection.\arabic{equation}}

\newcommand{\eq}{\begin{equation}}
\newcommand{\en}{\end{equation}}

\newcommand{\eqa}{\begin{eqnarray}}
\newcommand{\ena}{\end{eqnarray}}
\newcommand{\eqan}{\begin{eqnarray*}}
\newcommand{\enan}{\end{eqnarray*}}
\newcommand{\bear}[1]{\begin{array}{#1}}
\newcommand{\enar}{\end{array}}

\newcommand{\lbl}[1]{ \ifnum \DRAFT = 0
                             {\label {#1}}
                      \else  {\makebox[0in]{\raisebox{-2ex}{\tiny #1}
                                            \hspace{-6ex}}}
                             {\label {#1}}
                      \fi}
\newcommand{\rf}[1]{  \ifnum \DRAFT = 0
                             \ref {#1}
                      \else {\ref {#1}}
                            \makebox[0in]{\raisebox{-2ex}{\tiny #1}}
                      \fi}
\newcommand{\ct}[1]{  \ifnum \DRAFT = 0
                             \cite {#1}
                      \else {\cite {#1}}
                            \makebox[0in]{\raisebox{-2ex}{\tiny #1}}
                      \fi}

\newlength{\superigor}

\def \eroinappendice{0}
\newcounter{temporaneo}
\newcounter{appendice}

\newcommand{\sect}[1]{
\ifnum \eroinappendice =1
       \def\eroinappendice{0}
       \setcounter{appendice}{\value{section}}
       \setcounter{section}{\value{temporaneo}}
       \renewcommand{\theequation}{\thesection.\arabic{equation}}
       \fi
\setcounter{equation}{0}\section{#1}}

\newcommand{\app}[1]{
\ifnum \eroinappendice =0
       \def\eroinappendice{1}
       \setcounter{temporaneo}{\value{section}}
       \setcounter{section}{\value{appendice}}
       \renewcommand{\thesection}{Appendix \Alph{section}: }
       \renewcommand{\theequation}{\Alph{section}.\arabic{equation}}
      \fi
\setcounter{equation}{0}\section{#1}}

\newcommand{\draft}{ \ifnum \DRAFT =0  {} \else {\\ DRAFT} \fi}

\newcommand{\NP}[1]{Nucl. Phys.\ {\bf #1}\ }

\newcommand{\PL}[1]{Phys. Lett.\ {\bf #1}\ }
\newcommand{\PR}[1]{Phys. Rev\ {\bf #1}\ }

\newcommand{\RMP}[1]{Rev. Mod. Phys.\ {\bf #1}\ }

\def\sqr#1#2{{\vcenter{\hrule height.#2pt
     \hbox{\vrule width.#2pt height#1pt \kern#1pt
        \vrule width.#2pt}
     \hrule height.#2pt}}}

\def\thinspace{\kern .16667em}

\def\Dir{\nabla\kern-2ex\Big{/}}
\def\dslash{\partial\kern-1.5ex\Big{/}}

\def\reali{{\hbox{\s@ l\kern-.5ex R}}}
\def\naturali{{\hbox{\s@ l\kern-.5ex N}}}
\def\interi{{\mathchoice
 {\hbox{Z\kern-1.5mm Z}}
 {\hbox{Z\kern-1.5mm Z}}
 {\hbox{{Z\kern-1.2mm Z}}}
 {\hbox{{Z\kern-1.2mm Z}}}  }}

\def\unity{{\hbox{\s@ 1\kern-.8mm l}}}
\def\uno{{\hbox{ 1\kern-.8mm l}}}
\def\part{\partial}

\def\rd{\sqrt{2}}
\def\um{{1\over2}}

\def\rarr{\rightarrow}

\def\dag{\dagger}

\def\CB{{\cal B}}

\def\aa{\alpha}
\def\bb{\beta}

\def\dd{\delta}
\def\DD{\Delta}
\def\ee{\epsilon}

\def\gg{\gamma}
\def\GG{\Gamma}

\def\LL{\Lambda}

\newcommand{\vett}[2]{\left(
                      \begin{array}{c}
                     {#1} \\
                     {#2}
                     \end{array}
                     \right)
                    }

\begin{document}

\begin{titlepage}

\begin{flushright}
NORDITA-95/42 P\\
June 1995\\
hep-th/yymmddd
\end{flushright}
\vspace*{0.5cm}

\begin{center}
{\bf
\begin{Large}
{\bf
EXACT RESULTS FOR THE SUPERSYMMETRIC $G_2$ GAUGE THEORIES
\\}
\end{Large}
}
\vspace*{1.5cm}
          {\large Igor Pesando}
          \footnote{E-mail PESANDO@NBIVAX.NBI.DK,
            pesando@hetws4.nbi.dk, 31890::I\_PESANDO}
           \footnote{Work supported by the EU grant ERB4001GT930141.}
         \\[.3cm]
          NORDITA\\
          Blegdamsvej 17, DK-2100 Copenhagen \O \\
          Denmark\\
\end{center}
\vspace*{0.7cm}
\begin{abstract}
{
We study the $N=1$ supersymmetric gauge theories with $N_f$ flavors of
quarks in the fundamental vector representation.
We find dynamically generated superpotentials, smooth quantum moduli
space, quantum moduli space with additional mesons, non trivial IR
fixed points.
}
\end{abstract}
\vfill
\end{titlepage}

\setcounter{footnote}{0}

\def\un{\underline}
\def\para{\parallel}
\def\nf{{N_f}}
\sect{Introduction and conclusion.}
In the recent period there has been a lot of interest in
supersymmetric gauge theories; this is due to the fact that certain
 quantities are often exactly calculable because they are
extremely constrained by holomorphy and symmetries (for a review
see ref. (\ct{Se08013}) ).
These exact results provide  insight in the strong coupling region
where a lot of new interesting phenomena take place.
Some of them are universal, other specific.

One feature, which seems to be universal, is the sequence of phases into
which the theory with massless quarks moves when the number of flavors
is increased (for a short review see ref.s (\ct{Se05077}) ).
First when there are few flavors the theory has a dynamically
generated superpotential without ground state,
then it moves into a confining phase which is eventually followed by
a free magnetic one. Increasing further the number of massless quarks the
theory changes phase to a non-Abelian Coulomb phase and finally it
stops being asymptotically free and therefore it moves into the free
electric phase.

A second feature, which is perhaps the most exciting one,
is the existence of a dual
description in terms of a magnetic theory in the deep IR for a certain
range of the parameter $\nf$, the number of flavors.
In almost all the examples so far known the dual theory has
a different gauge group, which nevertheless belongs to the
same series of the Lie classification of the original one
(\ct{Se11149},\ct{IS}, \ct{IP},\ct{Ku03086}).
An exception to this
pattern is represented by the $SU(2 k)$ theory with an antisymmetric tensor
and vector matter (\ct{Be}) whose dual is the product of two groups.
Nevertheless in all the known cases it is true that the dual of a
simple group with matter in the elementary vector representation
is in the same Lie series.

It is therefore interesting trying to understand whether these
features are preserved in the case of special groups.
In particular it
would be very interesting to know which is the dual of the special groups.
The purpose of this work is to examine the simplest of all the special
groups, i.e. $G_2$.
We find that at least the first feature is preserved while we have not
being able to show whether the second is maintained.

A summary of our results is as follows:
for $\nf\le3$ flavours of vector matter we find dynamically
generated superpotentials associated either
with gaugino condensation for $\nf<3$ or with  instantons for $\nf=3$ which
lift the classical vacuum degeneracy  and imply the non existence of a
vacuum.
For $\nf>3$ there is still a quantum moduli space. In a generic point
of which the group is completely broken by the Higgs
mechanism. Classically there are singular submanifold where there is
an enhanced symmetry because some $W$ bosons are massless on them.
Quantistically for $\nf=4$ the moduli space is completely smooth,
without any singularity, because of instanton effect: the origin
does not belong to it. The theory confines.
For $\nf=5$ the quantum moduli space is the same of the classical one
but the singularity at the origin implies the existence of new mesons
rather than massless $W$; again at the origin the theory confines.
For $6\le \nf<12$ the theory flows to a non trivial interacting
superconformal field theory, which implies
a non-Abelian Coulomb phase.

\sect{The classical moduli space}
We will study the $N=1$ supersymmetric QCD with gauge group $G_2$ and
$\nf$ flavours of quarks $Q^{c f}$ in the (real) fundamental $\un{7}$
representation ($c=1\dots 7$, $f=1\dots \nf$).
The Wilsonian (one loop) beta function is
\eq
\bb_W=-{g^3\over 16 \pi^3}(12- \nf)
\en
and therefore the theory is asymptotically free for $\nf<12$.
The global symmetries are
\eq
\bear{|c|c|c||c|}\hline
    & SU(N_f) & U(1)_R            & U(1)_{Q^{f_0}} \\ \hline
Q^f &   N_f   &  {N_f-4\over N_f} & \dd_{f f_0}    \\ \hline
W   &    1    &         1         &    0           \\ \hline
\LL_{N_f}^{12-N_f}&1&   0         &    2           \\ \hline
\enar
\lbl{symm}
\en
where the last $U(1)_{Q^{f_0}}$ is anomalous. The theory also has a discrete
$Z_{2N_f}$ that acts as $Q\rarr e^{i\pi \over 2 N_f} Q$ which is
the part of $U(1)_A$ left unbroken by the anomaly.

Also in the case of $G_2$ it is possible to obtain the explicit form
of the classical moduli space; using the
explicit realization of the $G_2$ generators given in the appendix we find
(all the parameters are real)
\eq
\para Q_c^{f} \para =
\left(
\bear {c c c c c c c}
b_3\,{e^{{i\over 2}\,\left(2\,\phi_{16}+\phi_{31}-\phi_{34}+\pi\right) }}
  & 0 &a_3\,{e^{i\, \phi_{31}}}& 0& 0 &\dots& 0\\
b_4{e^{i\,\left(\phi_{16}+{{ \phi_{42}}\over 2}-{{\phi_{43}}\over 2}\right)}}
  & 0 & 0 &a_4\,{e^{i\, \phi_{42}}}& 0 &\dots& 0\\
b_4{e^{i\,\left(\phi_{16}-{{\phi_{42}}\over 2}+{{\phi_{43}}\over 2} +
         \pi  \right) }}
  & 0 & 0 &a_4\,{e^{i\, \phi_{43}}}& 0 &\dots& 0\\
b_3\,{e^{{i\over 2}\,\left( 2\,\phi_{16}-\phi_{31}+ \phi_{34} +
         3\,\pi  \right) }}
  & 0 &a_3\,{e^{i\, \phi_{34}}}&0 & 0&\dots& 0 \\
b_2{e^{i\,\left(  \phi_{16} + {{\phi_{25}}\over 2}-{{ \phi_{27}}\over 2}
          \right) }}
  &a_2\,{e^{i\, \phi_{25}}}& 0 &0 & 0&\dots& 0\\
a_1\,{e^{i\, \phi_{16}}}& 0& 0 &0 & 0&\dots& 0\\
b_2{e^{i\,\left(\phi_{16}-{{ \phi_{25}}\over 2}+{{ \phi_{27}}\over 2} +
         \pi  \right) }}
  &a_2\,{e^{i\, \phi_{27}}}& 0 & 0 & 0 &\dots& 0
\enar
\right)
\lbl{vacuum}
\en
along these flat directions there is the following pattern of symmetry
breaking $G_2 \stackrel{a_1\ne 0}\rarr SU(3) \stackrel{a_2\ne 0}\rarr
SU(2) \stackrel{a_3\ne 0}\rarr\emptyset$ and the generators left
unbroken are $\{ H,E_\aa\}\stackrel{a_1\ne 0}\rarr
\{H_1, H_2, E_{\pm 2}, E_{\pm 4}, E_{\pm 6} \} \stackrel{a_2\ne 0}\rarr
\{ H_2, E_{\pm 4}\}$.
It turns out examining the gauge invariant description of the moduli
space eq.s (\rf{mod-space}) that
the sequence in which we switch on the v.e.v is
$a_1,a_2,a_3,b_4,a_4,b_1,b_2,b_3$.

We can now write down the gauge invariant operators using the
two invariant tensors $g_{c d}$ and $\GG_{c_1 c_2 c_3}$
\eqa
M^{f g}&=&M^{g f}=g_{c d} Q^{c f} Q^{d g}
\nonumber\\
B^{f_1 f_2 f_3} &=&  \GG_{c_1 c_2 c_3}
                    Q^{c_1 f_1} Q^{c_2 f_2} Q^{c_3 f_3} ~~~\nf \ge 3
\nonumber\\
\CB^{f_1 f_2 f_3 f_4} &=& \GG^p_{. [c_1 c_2} \GG_{c_3 c_4] p}
                Q^{c_1 f_1} Q^{c_2 f_2} Q^{c_3 f_3} Q^{c_4 f_4} ~~~\nf \ge 4
\lbl{invariants}
\ena
with their charges under the global symmetries
\eq
\bear{|c|c|c||c|}\hline
              & SU(N_f)            & U(1)_R          &U(1)_{Q^{f_0}}\\ \hline
M^{f g}       & {\nf(\nf+1)\over 2}&2{N_f-4\over N_f}&
                                              \dd_{ff_0}+\dd_{g f_0}\\\hline
B^{f_1f_2f_3} &\vett{\nf}{3}&3{N_f-4\over N_f}&\sum_i \dd_{f_i f_0} \\ \hline
\CB^{f_1..f_4}& \vett{\nf}{4}      &4{N_f-4\over N_f}&   1          \\ \hline
\enar
\en
and then use the Bose statistic of $Q^{ c f}$ and the proprieties of
$\GG$ to deduce relation among these objects.

In particular the fact that we can only use up to two $\GG$ follows
from the reducibility of the product of three or more $\GG$ (as shown
in appendix eq. (\rf{red1})
and the explicit antisymmetrization over the flavour indices
in the last equation of eq. (\rf{invariants}), which implies
the constraint $\nf \ge 4$, is a consequence of
\eq
\CB^{f_1 f_2 | f_3 f_4}
=\CB^{[f_1 f_2 f_3 f_4]}+2 M^{f_1 [ f_3} M^{ f_4] f_2}
\lbl{red2}
\en

The gauge invariant description of the moduli space (\rf{vacuum}) is
given by
\eqa
M&=&\left(
\bear{c c c c }
-2\,{{a_{4}}^2}\,{e^{i\,\left(  \phi_{42} +  \phi_{43} \right) }} &
  0 & 0 & 0 \cr
 0 & -2\,{{a_{2}}^2}\,
   {e^{i\,\left(  \phi_{25} +  \phi_{27} \right) }} & 0 & 0 \cr
 0 & 0 &
  2\,{{a_{3}}^2}\,{e^{i\,\left(  \phi_{31} +  \phi_{34} \right) }} & 0 \cr
 0 & 0 & 0 & {e^{2\,i\, \phi_{16}}}\,
   \left( {{a_{1}}^2} + 2\sum_i b_i^2
  \right)  \cr
\enar
\right)
\nonumber\\
B^{1 2 3 }&=&
 i\,{\sqrt{2}}\,a_{2}\,a_{3}\,b_4
     {e^{{{-i}\over 2}\,\left(-2\,\phi_{16}+\phi_{42}+\phi_{43}\right)}}
      \,\left( {e^{i\,\left(\phi_{25}+\phi_{34}+\phi_{42}\right) }} +
       {e^{i\,\left(\phi_{27}+\phi_{31}+\phi_{43}\right) }} \right) \,
\nonumber\\
B^{1 2 4}&=&-
    {\sqrt{2}}\,a_{2}\,a_{4}\,b_3
    {e^{{{-i}\over 2}\,\left( -2\,\phi_{16}+\phi_{31}+\phi_{34}\right)}}
      \,\left( {e^{i\,\left(\phi_{25}+\phi_{34}+\phi_{42} \right) }} +
       {e^{i\,\left(\phi_{27}+\phi_{31}+\phi_{43} \right) }} \right)
\nonumber\\
B^{1 3 4}&=&
    i\,{\sqrt{2}}\,a_{3}\,a_{4}\, b_2
     {e^{{{-i}\over 2}\,\left(-2\,\phi_{16}+\phi_{25}+\phi_{27}\right)}}
      \,\left( {e^{i\,\left(\phi_{25}+\phi_{34}+\phi_{42} \right) }} +
       {e^{i\,\left(\phi_{27}+\phi_{31}+\phi_{43} \right) }} \right) \,
\nonumber\\
B^{2 3 4}&=&
i\,{\sqrt{2}}\,a_{2}\,a_{3}\,a_{4}\,
\left(
      {e^{i\,\left(  \phi_{25} +  \phi_{34} +  \phi_{42} \right) }} -
      {e^{i\,\left(  \phi_{27} +  \phi_{31} +  \phi_{43} \right) }}
\right)
\nonumber\\
\CB^{1 2 3 4}&=&
 {\sqrt{2}}\,a_{1}\,a_{2}\,a_{3}\,a_{4}\,
    {e^{i\,\phi_{16}}}\,\left( {e^
        {i\,\left(\phi_{25}+\phi_{34} +\phi_{42} \right) }} +
      {e^{i\,\left(  \phi_{27}+\phi_{31}+\phi_{43} \right) }} \right)
\lbl{mod-space}
\ena

\sect {$N_f=1,2$. A dynamically generated superpotential by gaugino
condensation with no vacuum.}

Using the symmetries  eq. (\rf{symm}) the only dynamical
generated superpotential, which is allowed, is
\eq
W_{N_f}= A_{N_f}\left( \LL_{N_f}^{12-N_f}\over \det M\right)
^{1\over 4-N_f}
\lbl{w12}
\en
Adding $W_{\mbox{tree}}= m M$ and integrating away the massive
quarks we find $A_{N_f}=(4-N_f)\left(A_0\over 4\right)^{4\over
4-N_f}$.
We fix $A_0=4$ in such a way that the pure $G_2$ YM theory has the effective
superpotential $W_0= 4(\LL_{N_f}^{12-N_f} \det m )^{1\over 4}$
which shows that the pure $G_2$ SYM has 4 different vacua.

As in ref. (\ct{ADS}) this superpotential is generated by gaugino
condensation in the $SU(4-\nf)$ YM theory left unbroken by $<Q>$.
This can be checked along the flat direction with $<Q^{c 1}>=\dd^{c 6} a_1$
of the $\nf=2$ theory where the low energy theory left by the Higgs
mechanism is $SU(3)$ with $\nf=1$ flavour: we find that the two scales
are relate by $\LL^8_{3,1}={\LL^{10}_2\over 2 a_1^2}$ since $M^{2 2}=2
{\hat M}^{2|2}$ where ${\hat M}^{2|2}$ is the gauge invariant meson of
the $SU(3)$ theory.
In the $SU(3)$ SQCD obtained higgsing $G_2$
the $\bar {\un{3}}$ quarks transform naturally in the $\nf$
representation of the
global gauge group $SU(\nf)\times SU(\nf) \times U(1)_V \times U(1)_R$
exactly as the $\un{3}$ do but differently from the usual assignment.

\sect{$N_f=3$. An instanton generated superpotential with no vacuum.}

The superpotential eq. (\rf{w12}) also makes sense for
$N_f=3$ and it reduces nicely to $W_{2}$ when adding
$W_{\mbox{tree}}=m_{3 3} M^{3 3}$ and integrating the third massive quark away
but when $N_f=3$ there is another gauge invariant available besides
$M^{f g}$ (eq. (\rf{invariants})) and it is $B\equiv B^{1 2 3}$

The classical vacua can be described with these 7 gauge invariant
operators without any constraints as the naive counting of the real
d.o.f shows: $ \sharp(Q)-2 dim(G_2)=\sharp(M)+\sharp(B)=14$.
This can also directly be confirmed with the tensor analysis and by
the direct inspection of the classical moduli space (\rf{mod-space}).

Using the symmetries eq. (\rf{symm}) we can write the most general
superpotential as
\eq
W_3={\LL^9_3\over \det M} f\left( {B^2\over \det M}\right)
\en
where $f(u)$ is an arbitrary function.
Adding $W_{\mbox{tree}}=m M+ b B$ we get
\eq
W_{\mbox{eff}}={\LL^9_3\over \det M}{\bar f}\left( {B^2\over \det M},
{\det m (\det M)\over (m M)^3},
{m M~ \det M \over \LL^9_3},
{b B~ \det M \over \LL^9_3} \right)
\en
Since for $\det M\ne 0$ the group is completely broken we expect
instanton corrections which contribute as $(\LL^9_3)^n$ with $n\ge0$,
i.e. ${\bar f}(x_1,x_2,x_3,x_4)=\sum_{n+m\ge -1}{\bar f}_{n
m}(x_1,x_2) x_3^n x_4^m$.
The further necessities to have a smooth $m=0$ and $b=0$ limit imply
\eq
W_{\mbox{eff}}={\LL^9_3\over \det M}f\left( {B^2\over \det M}\right)
+m M+b B
\en
In order to determine $f$ we use the integrate in technique (\ct{In}) (whose
notation we use) since the Principle of Linearity is satisfied.
Integrating away $Q^{c 3}$ at the classical level in
$W_{\mbox{tree}}$ we get $W_{\mbox{tree, d}}=-{b^2\over 4 m_{3 3}}\det{\hat M}$
where ${\hat M}$ are the gauge invariant mesons describing the $\nf=2$
down theory.
The most general $W_\DD$ is
$W_\DD=\left( {\LL^{10}_2\over \det{\hat M}}\right)^\um w \left({b^2
(\det{\hat M})^3\over\LL^{10}_2 m_{3 3}^2 }\right)$.
As  explained in ref. (\ct{In}) we must have $W_\DD|_{b=0}=0$,
$\lim_{m_{3 3}\rarr\infty} W_\DD =0$ and
$\lim_{\LL^{10}_2\rarr0}W_\DD=0$
(because switching off the gauge interaction we are left only with
$W_{\mbox{tree, d}}$)
Tuning the limits
$m_{3 3}\rarr\infty$  and $\LL^{10}_2\rarr0$ we get $W_\DD=0$.
Assuming simple thresholds,
we can now integrate away $b$ and $m_{3 3}$ from the expression
$W_n=\left( {\LL^{10}_2\over \det{\hat M}}\right)^\um
-{b^2\over 4 m_{3 3}} \det {\hat M}- m_{3 3}{  M}^{3 3}- b{ B}$.
We  obtain
\eq
W_3={\LL^9_3\over \det { M} - { B}^2}
\lbl{w3}
\en
where  ${f}({ u})={1\over 1-{ u}}$.
An independent test of this superpotential can be obtained higgsing
the theory to $SU(3)$ by $<Q^{c 1}>=\dd^{c 6} a_1$.
We find ($f,g\ge 2$) $M^{f g}={\hat M}^{f|g}+{\hat M}^{g|f}$
and $B=i a_1 ( {\hat M}^{2|3}-{\hat M}^{3|2})$ which gives the usual
superpotential for $SU(3)$ with $\nf=2$ when we identify the two
scales as $\LL^7_{3,2}={\LL^9_3\over 4 a_1^2}$.

Because of this superpotential the theory has not a vacuum and exhibits
the runaway phenomenon.

\sect{$\nf=4$. A smooth quantum moduli space.}

In this case because the quarks are not charged under the $U(1)_R$
symmetry, it is not possible to generate a superpotential, which has
$U(1)_R$ charge two.
Nevertheless the classical theory can be described with the 15 gauge invariants
$M^{ f g}$, $B_f \equiv {1\over 3!} \ee_{f f_1 f_2 f_3} B^{f_1 f_2
f_3}$ and $\CB \equiv \CB^{1 2 3 4}$ and one constraint
\eq
\det M - \CB^2 - B_f M^{ f g} B_g=0
\lbl{cl-co-4}
\en
The necessity for this constraint is easily seen with the help of the
naive counting of the complex d.o.f $\sharp(Q)-\sharp(G_2)=14$ and it
can be obtained with the help of the tensor analysis and using the
Bose symmetry of the $Q$s.

If we turn on $W_{\mbox{tree}}= m M$ and we use the symmetries, we find that
\eq
<M^{f g}>= k_1 \LL_4^2 (\det m)^{1\over 4} (m^{-1})^{f g}
{}~~~
<B_f>=0
{}~~~
<\CB>= k_2 \LL_4^4
\lbl{vev-4}
\en
with some unknown constants $k_1,k_2$. At the quantum level the constraint
eq. (\rf{cl-co-4}) needs to be modified to
\eq
\det M - \CB^2 - B_f M^{ f g} B_g=(k_1^4- k_2^2)\LL_4^8
\lbl{q-co-4}
\en
in order to be able to accommodate the v.e.v. eq. (\rf{vev-4}).
As in ref. (\ct{Se02044}) this modification is a pure one instanton effect
and it has as a consequence the smoothing out of the classical
singularity at the origin.
One could wonder whether it can happen that the two contributions to
the r.h.s. of eq. (\rf{q-co-4}) cancel each other. If so there would be
an enhanced symmetry at the origin of the moduli space since there
would be no smoothing of the singularity.
The answer is that it cannot happen
as it can be seen decoupling one flavour from the $\nf=5$ theory where
this possibility does not exist.
Moreover turning on $W_{\mbox{tree}}=m_{4 4} M^{ 4 4}$ and decoupling
the fourth quark we recover the superpotential for $\nf=3$
eq. (\rf{w3}) when we identify $\LL^9_{3,3}= m_{4 4}(k_1^4- k_2^2)\LL^8_4$.
It should therefore be possible using the $\overline{DR}$ scheme
to set $(k_1^4- k_2^2)=1$.

Another independent check of this result is obtained by higgsing the
theory to $SU(3)$ by $<Q^{c 1}>=\dd^{c 6} a_1$. We
get $B^{2 3 4}=i\rd ({\hat B}-{\hat{\tilde B}})$,
$B^{1 f g}=i a_1( {\hat M}^{f|g}-{\hat M}^{g|f})$,
$\CB=a_1 \rd ({\hat B}+{\hat{\tilde B}})$
which imply the usual $\nf=N_c=3$ constraint for the $SU(3)$ theory
(\ct{Se02044}) when we identify
$\LL^6_{3,3}=(k_1^4-k_2^2) {\LL^{8}_4\over 8 a_1^2}$.

In order to verify the consistency of previous picture we can check the 't
Hooft anomaly matching condition (\ct{tH}) at the point $M^{f g}=B_f=0$ and
$\CB= \sqrt{-(k_1^4-k_2^2)} \LL^4_4$.
At this point the whole global symmetry is unbroken. The microscopic
 fermions transform
in the $14\times (\un{1})_{1}\oplus 7\times(\un{\nf})_{-{4\over\nf}} $ of
$SU(\nf=4)\times U(1)_R$ while the macroscopic ones in the
$\left({ \nf(\nf+1)\over 2}\right)_{-1}\oplus ({\bar \nf})_{-1}$.
This gives the following identities for the 't Hooft anomalies (we use
the coefficient computed in ref. (\ct{tH}) p. 153)
\eqa
U(1)_R && -14= -{ \nf(\nf+1)\over 2}-\nf
\nonumber\\
U(1)_R^3 && 14-{7\cdot 64\over\nf^2}= -{ \nf(\nf+1)\over 2}-\nf
\nonumber\\
SU(\nf)^2 U(1)_R && -7 d_2(\nf)=
 -d_2\left({\nf(\nf+1)\over2}\right)-d_2({\bar \nf})
\nonumber\\
SU(\nf)^3 && 7  d_3(\nf)= d_3\left({\nf(\nf+1)\over 2}\right)+d_3({\bar \nf})
\ena
where $d_2(\un{r})$ and $d_3(\un{r})$ are respectively the second and
third Casimir of the irrep $\un{r}$.

\sect{$\nf=5$. Confinement without chiral symmetry breaking.}
Similarly to
the case $\nf=4$ we find that the classical moduli space is
described by the gauge invariants given in eq. (\rf{invariants}) with
3 constraints
\eqa
&&
\ee_{f f_1\dots f_4}\ee_{g g_1\dots g_4}
M^{f_1 g_1}\dots M^{f_4 g_4}
-\CB_{f} \CB_{g}
-  B_{f f_1} B_{g g_1} M^{f_1 g_1} =0
\nonumber\\
&&
M^{f g} \CB_g
+{1\over 8} \ee^{f f_1\dots f_4} B_{f_1 f_2} B_{f_3 f_4}=0
\nonumber\\
&&
B_{f_1 g_1} M^{f f_1} M^{g g_1}
+\um \ee^{ f g f_1 f_2 f_3} \CB_{f_1} B_{f_2 f_3}=0
\ena
where $B_{f g}={1\over 3!}  \ee_{ f g f_1 f_2 f_3}  B^{f_1 f_2 f_3}$
and $\CB_f= {1\over 4!}  \ee_{ f f_1 f_2 f_3 f_4} \CB^{f_1 f_2 f_3 f_4}$.

Turning on  $W_{\mbox{tree}}= m M$ we find that the symmetries imply that
$<M^{f g}>= k_1 \LL_5^2 (\det m)^{1\over 4} (m^{-1})^{f g}$,
$<B^{f_1 f_2 f_3}>=0$ and
$<\CB^{f_1\dots f_4}>=0$ and we must therefore modify the classical
constraint to
\eq
\ee_{f f_1\dots f_4}\ee_{g g_1\dots g_4}
M^{f_1 g_1}\dots M^{f_4 g_4}
-\CB_{f} \CB_{g}
-  B_{f f_1} B_{g g_1} M^{f_1 g_1}
= k_1^4 \LL_5^7 m_{f g}
\en
Following the line of thought of ref. (\ct{Se02044}) we are lead to
conclude that all the gauge invariant fields are required to a
complete description of the quantum theory since we can completely
fill a neighbourhood of the origin of the quantum moduli space tuning
the ``external'' sources in $W_{\mbox{tree}}=m M+ b B+ c\CB$, i.e. we
find all the possible values of $M$, $B$ and $\CB$ for $m,b,c\ne0$.

This satisfies a highly non trivial check of consistency: the 't Hooft
anomaly consistency conditions.
At the origin of the quantum moduli space the whole global symmetry
$SU(5)\times U(1)_R$ is unbroken and the microscopic fields transform
in $ 14\times (\un{1})_{1}\oplus 7\times(\un{5})_{-{4\over5}}$ while
the macroscopic ones in the
$(\un{15})_{-{3\over 5}}\oplus (\un{10})_{-{2\over 5}}\oplus
(\un{5})_{-{1\over 5}}$.
The 't Hooft anomaly conditions, which are satisfied, are
\eqa
U(1)_R && -14= 15\cdot\left(-{3\over 5} \right)+10\cdot\left(-{2\over 5}\right)
              +5\cdot\left(-{1\over 5} \right)
\nonumber\\
U(1)_R^3 &&
14-{64\cdot7\over\nf^2}= 15\cdot\left(-{3\over 5} \right)^3
 +10\cdot\left(-{2\over 5}\right)^3
 +5\cdot\left(-{1\over 5} \right)^3
\nonumber\\
SU(\nf)^2 U(1)_R && -{28\over \nf} d_2(\nf)=
 d_2(15)\cdot\left(-{3\over 5} \right)
+d_2(10)\cdot\left(-{2\over 5} \right)
+d_2(5)\cdot\left(-{1\over 5} \right)
\nonumber\\
SU(\nf)^3 && 7  d_3(\nf)=  d_3(15)+d_3(10)+ d_3(5)
\ena

Since the theory makes sense at the origin too and we can describe it
using all the gauge invariant fields we have, we can expect to be able
to find a unique superpotential for the effective low energy
Lagrangian.
The unique superpotential which respects the symmetries, reproduces
the flat directions correctly and yields the previous superpotentials
when integrating out quarks is given by
\eq
W_5=
{1\over \LL^7_5}
\left(-det M
+\um B_{f_1 f_2} B_{g_1 g_2} M^{f_1 g_1} M^{f_2 g_2}
+\CB_f \CB_g M^{f g}
+{1\over 4} \ee^{f_1\dots f_5}\CB_{f_1} B_{f_2 f_3} B_{f_4 f_5}
\right)
\en

\sect{$\nf\ge 6$. The interacting superconformal field theory. }

As in the previous case the quantum moduli space has a singularity at
the origin but
now we cannot satisfy the 't Hooft conditions.  The simplest
one, the $U(1)_R$ condition, tells that there are not enough fermions with
negative $U(1)_R$ charge ( for the $\nf=6$ we get $-14$ from the
microscopic point of view while the macroscopic contribution is $-2$)
but we cannot construct any further operator whose fermions have  negative
$U(1)_R$ charge.
The theory at the origin should be in a non Abelian Coulomb phase
since it can be higgsed to a $SU(3)$ theory with $\nf\ge 5$ which is
known to be either in such a phase ($\nf\le 8$) or not asymptotically
free ($\nf\ge 9$).
Let us therefore examine the dimension of the gauge invariant
operators which we have given in eq. (\rf{invariants}), in the deep
infrared supposing that the theory is described by a $N=1$
superconformal theory (\ct{Se11149}), we get
\eqa
D(M^{f g})&=& {3\over 2} D(M^{f g})=3-{12\over \nf}
\nonumber\\
D(B^{f_1 f_2 f_3})&=&{9\over 2}-{18\over \nf}
\nonumber\\
D(\CB^{f_1 f_2 f_3 f_4})&=&6-{24\over \nf}
\ena
{}From the theory of representation of $N=1$ superconformal theory we
know that a unitary representation necessary has $D\ge1$ for all the
operators, this implies $\nf\ge6$.
In particular for $\nf=6$ if the theory is superconformal in the IR,
the mesonic fields $M^{f g}$ is free.
The non existence of a gap between the confining phase and the
non Abelian Coulomb phase is nothing peculiar of this theory since it
happens for $SU(2)$ and $SU(3)$ theories with fundamental matter too.

We could wonder whether it is now necessary to have a dual theory
since the very reason  calling for a dual
theory in the $U(n)$ (\ct{Se11149}), $SO(n)$ (\ct{IS}) and $Sp(n)$
(\ct{IP}) gauge theories with
fundamental matter, i.e. the non existence of a unitary theory in term of the
original gauge invariant fields in the deep IR, has disappeared and
everything proceeds
smoothly from $\nf=0$ to $\nf=12$ where the theory ceases to be
asymptotically free.
The answer is yes, we need to have a dual description because $G_2$
with $\nf=6$ can be obtained higgsing $SO(7)$ with $\nf=7$ flavours of
matter in the spinorial irrep $\un{8}$ and this theory should be in
the non-Abelian Coulomb
phase since $G_2$ is but on the other side its meson does not belong
to a unitary representation of a superconformal theory.

We can therefore try to construct the dual theory explicitly.
To this purpose we want to make use of the commutativity of the
diagram
\eq
\bear{l c r}
(G_2, \nf\cdot \un{7})
 & \stackrel{\mbox{\tiny dual}}{\Longrightarrow}
 & ({\tilde G}_2(\nf), \nf\cdot(\sum_i \un{r_i})
    \oplus \mbox{gauge invariants}) \cr
\mbox{\tiny Higgs}\downarrow
 &
 & \downarrow \mbox{\tiny decoupling}\cr
 (SU(3), (\nf-1)\cdot(\un{3}+\bar{\un{3}})  &
  \stackrel{\mbox{\tiny dual}}{\Longrightarrow} &
    (SU(\nf-4),(\nf-1)\cdot (\un{\nf-4}+\un{\overline{\nf-4}} )\oplus
    {{\hat M}}^{{\hat f}|{\hat g}})\cr
\enar
\en
where we are assuming that the decoupling is the corresponding
phenomenon in the magnetic theory of the higgsing in the electric
theory and that there is a unique dual to the $SU(3)$ gauge theory
with vector matter. With this assumption we can conclude that ${\tilde
G}_2(\nf)=SU(\nf-4)$ and $\sum_i \un{r_i}=\un{\nf-4}+{\overline{\un{\nf-4}}}$.
Now we run into troubles since the global symmetries of the $SU(3)$ and
$G_2$ theories are quite different: the $SU(3)$ symmetries
$SU(\nf-1)\times SU(\nf-1)\times U(1)_V \times U(1)_R$ are not
a subgroup of the $G_2$ symmetries $SU(\nf)\times U(1)_R$. We can
partially solve the problem with the superpotential
${\tilde W}=M^{f g} q_f {\bar q}_g$
which breaks $SU(\nf)\times SU(\nf)$ to $SU(\nf)$
since $M^{f g}$ is symmetric.
We are then obliged to introduce  $B^{f g h}$ as elementary
field of the dual since we want to identify $B^{1 f g}$ with the antisymmetric
part of ${{\hat M}}^{{\hat f}|{\hat g}}$. Another reason to introduce
$B$ as ane elemntary fiels is that it is not possible
to construct any operator using the dual quarks with the same $U(1)_R$
charge of $B$.
But we are now faced with the problem of eliminating a linear
combination of the dual baryon and dual antibaryon from the chiral
ring since we have only one baryon with four indeces in the electric
theory. Associated with this problem there is also the way of
eliminating the global exceeding $U(1)_V$ symmetry.
We were not able to solve this problem but we hope to return on it in
another paper.

\app{The group $G_2$}
The group $G_2$ is the group to which $SO(7)$ is spontaneously broken
by a spinor $S$ in the $\underline{8}$ of $SO(7)$.
$G_2$ has therefore dimension 14 and rank 2 and all its irreps are real.

The branching rules of $SO(7)$ in $G_2$ are
\eqa
\un{7}&\rarr& \un{7}
\nonumber\\
\un{21}&\rarr& \un{7} +\un{14}
\ena
and both $\un{7}$ and $\un{14}$ are elementary irreps of $G_2$ but
only $\un{7}$ is a simple irrep.

In particular
$\un{7}\times\un{7}=(\un{1}+\un{27})_{\mbox{symm}}
                   +(\un{7}+\un{14})_{\mbox{antisymm}}
$
therefore the confining phase and the Higgs phase are
indistinguishable since there is no Wilson loop which cannot be
shielded  by quarks.

$G_2$ is characterised by two invariants (\ct{BDFL})
\begin{enumerate}
\item $\dd_{a b}$
\item $\GG_{a b c}$ which is totally antisymmetric
\end{enumerate}
where $a,b\dots =1,\dots 7$ and
\eq
\GG_{i..}^{.l n} \GG_{j.n}^{.k.}+\GG_{j..}^{.l n} \GG_{i.n}^{.k.}
= 2\dd_{i j} \dd^{l k} - 2\dd^l_{\{ i} \dd_{j \}}^k
= 2\dd_{i j} \dd^{l k} - \dd_{i}^{l} \dd_{j}^{k} -\dd_{i}^{k} \dd_{j}^{l}
\lbl{red0}
\en

Moreover $\GG_{a b c}$ can be interpreted
as $\GG_{a b c}=<{\bar S}\gg_{a b c}S>$ in view of the embedding
$G_2\subset SO(7)$ described at the beginning.

The previous formula eq. (\rf{red0}) allows the complete reduction of
the products of three or more $\GG$:
\eq
\GG^{m n s} \GG_{s t i} \GG^t_{. j k}=
2 \dd_{i [j } \GG_{k]}^{m n} - 6 \dd^{[ m}_{[ i} \GG^{\ . \ . n]}_{ j k ]}
\lbl{red1}
\en

In order to be able to find the constraints among the gauge invariant
fields we need the following decompositions into irreducible parts
\eqa
T_{[c_1 c_2 c_3]}&=&
{1\over7\cdot 3!} \GG_{c_1 c_2 c_3}  T
+{1\over 24}  \GG^p_{. [c_1 c_2} \GG_{c_3] k p} T^k
+{3\over 4} \left( \GG_{ . [c_1 c_2}^{\{k} \dd_{c_3]}^{l\}}
-{1\over 7} \GG_{c_1 c_2 c_3} g^{k l}\right) T_{\{ k l\}}
\nonumber\\
&=&
-{1\over 12} \GG_{c_1 c_2 c_3} (\GG^{d_1 d_2 d_3} T_{d_1 d_2 d_3})
+{1\over 2} (\GG^2)^{ k l}_{[c_1 c_2} T_{c_3] k l}
+{1\over4 }  \GG_{ d_1 [c_1 c_2} \GG_{c_3] d_2 d_3} T^{d_1 d_2 d_3}
\nonumber\\
{}~
\ena
where $(\GG^2)^{ k l}_{c_1 c_2}=\GG^{s k l} \GG_{s c_1 c_2}$ and
$T$, $T_k$ and $T_{\{ k l\}}$ are the $\un{1}$, $\un{7}$ and
$\un{27}$ irreducible parts given by
\eqa
T&=&\GG^{c_1 c_2 c_3} T_{c_1 c_2 c_3}
\nonumber\\
T_k&=& \GG^p_{. [c_1 c_2} \GG_{c_3] k p} T^{c_1 c_2 c_3}
\nonumber\\
T_{\{ k l\}} &=&  \left( \GG_{c_1 c_2 \{k } \dd_{ l\} c_3}
-{1\over 7} g_{k l}\GG_{c_1 c_2 c_3} \right) T^{c_1 c_2 c_3}
\ena

Moreover we need
\eq
T_{[c_1\dots c_4]}=
{1\over7\cdot 4!} \GG^p_{.[c_1 c_2} \GG_{c_3 c_4]p} T
+{1\over 6}  \GG_{[c_1 c_2 c_3} \dd_{c_4] k} T^k
+{3\over 16} \left( \GG_{k [c_1 c_2} \GG_{c_3 c_4] l}
-{1\over 7}\dd_{k l} \GG_{p [c_1 c_2} \GG_{c_3 c_4] p}\right) T_{\{ k l\}}
\en
where $T$, $T_k$ and $T_{\{ k l\}}$ are the $\un{1}$, $\un{7}$ and
$\un{27}$ irreducible parts given by
\eqa
T&=&\GG_{p[c_1 c_2} \GG_{c_3 c_4]p} T^{[c_1\dots c_4]}
\nonumber\\
T_k&=&\GG^{c_1 c_2 c_3} T_{[c_1 c_2 c_3 k]}
\nonumber\\
T_{\{ k l\}} &=& \left( \GG_{k [c_1 c_2} \GG_{c_3 c_4] l}
     -{1\over 7} g_{k l} \GG^p_{.[c_1 c_2} \GG_{c_3 c_4] p}\right)
        T_{[c_1\dots c_4]}
\ena

We also need
\eq
T_{[c_1\dots c_5]}=
{5\over72} \GG^p_{.[c_1 c_2} \GG_{c_3 c_4 c_5]} T_p
+{5\over27} \left( \um \GG_{[c_1 c_2 c_3} \GG_{c_4 c_5] p} \GG^{ p k l}
+ \GG_{[c_1 c_2 c_3}\GG_{c_4 . .}^{. k p}\GG_{c_5]. p}^{. l .}
\right) T_{[ k l]}^{(14)}
\en
where $T_k$ and $T_{[ k l]}^{(14)}$ are the $\un{1}$ and $\un{14}$
irreducible parts  given by
\eqa
T_k&=&\GG^{c_1 c_2 c_3} \GG_{c_4 c_5 k} T^{[c_1 c_2 c_3 c_4 c_5]}
\nonumber\\
T^{[ k l](14)}&=&
\left( \um \GG_{[c_1 c_2 c_3} \GG_{c_4 c_5] p} \GG^{ p k l}
+ \GG_{[c_1 c_2 c_3}\GG_{c_4 . .}^{. k p}\GG_{c_5]. p}^{. l .}
\right) T^{[c_1 c_2 c_3 c_4 c_5]}
\ena

\app{Explicit representation of $G_2$}

We took the explicit representation from the first reference of
(\ct{BDFL}) which we checked and corrected.
{

$$g_{ i j}=\left(
\matrix{ 0 & 0 & 0 & 1 & 0 & 0 & 0 \cr 0 & 0 & -1 & 0 & 0 & 0 & 0 \cr 0 & -1
   & 0 & 0 & 0 & 0 & 0 \cr 1 & 0 & 0 & 0 & 0 & 0 & 0 \cr 0 & 0 & 0 & 0 & 0 & 0
   & -1 \cr 0 & 0 & 0 & 0 & 0 & 1 & 0 \cr 0 & 0 & 0 & 0 & -1 & 0 & 0 \cr  }
\right)
$$

$$
H_1 =\left(
 \matrix{ {1\over {4\,{\sqrt{3}}}} & 0 & 0 & 0 & 0 & 0 & 0 \cr 0 &
  {{-1}\over {4\,{\sqrt{3}}}} & 0 & 0 & 0 & 0 & 0 \cr 0 & 0 &
  {1\over {4\,{\sqrt{3}}}} & 0 & 0 & 0 & 0 \cr 0 & 0 & 0 &
  {{-1}\over {4\,{\sqrt{3}}}} & 0 & 0 & 0 \cr 0 & 0 & 0 & 0 &
  {1\over {2\,{\sqrt{3}}}} & 0 & 0 \cr 0 & 0 & 0 & 0 & 0 & 0 & 0 \cr 0 & 0 & 0
   & 0 & 0 & 0 & {{-1}\over {2\,{\sqrt{3}}}} \cr  }
\right)
{}~
H_2 =\left(
 \matrix{ {1\over 4} & 0 & 0 & 0 & 0 & 0 & 0 \cr 0 & {1\over 4} & 0 & 0 & 0 & 0
   & 0 \cr 0 & 0 & -{1\over 4} & 0 & 0 & 0 & 0 \cr 0 & 0 & 0 & -{1\over 4} & 0
   & 0 & 0 \cr 0 & 0 & 0 & 0 & 0 & 0 & 0 \cr 0 & 0 & 0 & 0 & 0 & 0 & 0 \cr 0
   & 0 & 0 & 0 & 0 & 0 & 0 \cr  }
\right)
$$
$$
E_{1} =\left(
 \matrix{ 0 & {1\over {2\,{\sqrt{6}}}} & 0 & 0 & 0 & 0 & 0 \cr 0 & 0 & 0 & 0 &
  0 & 0 & 0 \cr 0 & 0 & 0 & {1\over {2\,{\sqrt{6}}}} & 0 & 0 & 0 \cr 0 & 0 & 0
   & 0 & 0 & 0 & 0 \cr 0 & 0 & 0 & 0 & 0 & {1\over {2\,{\sqrt{3}}}} & 0 \cr 0
   & 0 & 0 & 0 & 0 & 0 & {1\over {2\,{\sqrt{3}}}} \cr 0 & 0 & 0 & 0 & 0 & 0 &
  0 \cr  }
\right)
{}~
E_{2} =\left(
 \matrix{ 0 & 0 & 0 & 0 & 0 & 0 & {1\over {2\,{\sqrt{2}}}} \cr 0 & 0 & 0 & 0 &
  0 & 0 & 0 \cr 0 & 0 & 0 & 0 & 0 & 0 & 0 \cr 0 & 0 & 0 & 0 & 0 & 0 & 0 \cr 0
   & 0 & 0 & {1\over {2\,{\sqrt{2}}}} & 0 & 0 & 0 \cr 0 & 0 & 0 & 0 & 0 & 0 &
  0 \cr 0 & 0 & 0 & 0 & 0 & 0 & 0 \cr  }
\right)
$$
$$
E_{3} =\left(
 \matrix{ 0 & 0 & 0 & 0 & 0 & {1\over {2\,{\sqrt{3}}}} & 0 \cr 0 & 0 & 0 & 0 &
  0 & 0 & {{-1}\over {2\,{\sqrt{6}}}} \cr 0 & 0 & 0 & 0 & 0 & 0 & 0 \cr 0 & 0
   & 0 & 0 & 0 & 0 & 0 \cr 0 & 0 & {1\over {2\,{\sqrt{6}}}} & 0 & 0 & 0 & 0
   \cr 0 & 0 & 0 & {{-1}\over {2\,{\sqrt{3}}}} & 0 & 0 & 0 \cr 0 & 0 & 0 & 0
   & 0 & 0 & 0 \cr  }
\right)
{}~
E_{4} =\left(
 \matrix{ 0 & 0 & {1\over {2\,{\sqrt{2}}}} & 0 & 0 & 0 & 0 \cr 0 & 0 & 0 &
  {1\over {2\,{\sqrt{2}}}} & 0 & 0 & 0 \cr 0 & 0 & 0 & 0 & 0 & 0 & 0 \cr 0 & 0
   & 0 & 0 & 0 & 0 & 0 \cr 0 & 0 & 0 & 0 & 0 & 0 & 0 \cr 0 & 0 & 0 & 0 & 0 & 0
   & 0 \cr 0 & 0 & 0 & 0 & 0 & 0 & 0 \cr  }
\right)
$$
$$
E_{5} =\left(
 \matrix{ 0 & 0 & 0 & 0 & {{-1}\over {2\,{\sqrt{6}}}} & 0 & 0 \cr 0 & 0 & 0 & 0
   & 0 & {1\over {2\,{\sqrt{3}}}} & 0 \cr 0 & 0 & 0 & 0 & 0 & 0 & 0 \cr 0 & 0
   & 0 & 0 & 0 & 0 & 0 \cr 0 & 0 & 0 & 0 & 0 & 0 & 0 \cr 0 & 0 &
  {1\over {2\,{\sqrt{3}}}} & 0 & 0 & 0 & 0 \cr 0 & 0 & 0 &
  {{-1}\over {2\,{\sqrt{6}}}} & 0 & 0 & 0 \cr  }
\right)
{}~
E_{6} =\left(
 \matrix{ 0 & 0 & 0 & 0 & 0 & 0 & 0 \cr 0 & 0 & 0 & 0 &
  {{-1}\over {2\,{\sqrt{2}}}} & 0 & 0 \cr 0 & 0 & 0 & 0 & 0 & 0 & 0 \cr 0 & 0
   & 0 & 0 & 0 & 0 & 0 \cr 0 & 0 & 0 & 0 & 0 & 0 & 0 \cr 0 & 0 & 0 & 0 & 0 & 0
   & 0 \cr 0 & 0 & {1\over {2\,{\sqrt{2}}}} & 0 & 0 & 0 & 0 \cr  }
\right)
$$
$$ E_{-\aa}=E_\aa^\dag$$
$$ tr(E_{-\aa}E_\aa)=tr(H_i^2)={1\over 4}$$
}

$$N^4_{2,6}=N^2_{4,-6}=N^6_{-2,4}=N^3_{2,-1}=N^2_{3,1}=N^{-1}_{-2,3}$$
$$=N^1_{5,6}=N^5_{1,6}=N^6_{-1,5}=N^{-5}_{-4,3}=N^3_{4,-5}=N^{-4}_{-5,-3}
={1\over 2 \rd}$$
$$N^3_{1,5}=N^1_{3,-5}=N^5_{-1,3}={1\over \sqrt{6}}$$

$$\aa_{1}=({1\over 2\sqrt{3}},0) \;
\aa_{3}=({1\over 4\sqrt{3}},{1\over 4})
\aa_5=(-{1\over 4\sqrt{3}},{1\over 4}) \;
$$
$$
\aa_{2}=({\sqrt{3}\over4},{1\over 4}) \;
\aa_4=(0,\um) \;
\aa_6=(-{\sqrt{3}\over4},{1\over 4})
$$

$$
\GG^{1 3 7}=\GG^{2 4 5}= i\rd
{}~~~~
\GG^{1 4 6}=\GG^{2 3 6}= -\GG^{5 6 7}= -i
$$

\app{$SU(3)\subset G_2$}
We will consider the simplest way of embedding (higgsing) $SU(3)$ into $ G_2$.
This is achieved by turning on the v.e.v $<Q^c>=a \dd^c_6$.
In this way the  $SU(3)$ is generated by
$\{H_1, H_2, E_{\pm 2}, E_{\pm 4}, E_{\pm 6} \}$.
Using the matrices of the previous appendix is easy to realize that
\eqa
 \{ Q^1, Q^7, Q^3\}&\rarr& q^c
\nonumber\\
 \{ Q_1, Q_7, Q_3\}\equiv\{ Q^4, -Q^5, -Q^2\}&\rarr& {\bar q}_c
\nonumber\\
\{Q^6\}&\rarr& s
\lbl{branch}
\ena
where $q^c$, ${\bar q}_c$ and $s$ transform respectively as $\un{3}$,
$\overline{\un{3}}$ and $\un{1}$ of $SU(3)$.

Using the previous formulae eq.s (\rf{branch}) we can decompose the
gauge invariants operators (\rf{invariants}), we get ($f,g\ge 2$)
\def\mh{{\hat M}}
\def\bh{{\hat B}}
\def\bth{{\hat{\bar B}}}

\eqa
M^{f g}&=& \mh^{f g}+\mh^{g f}
\nonumber\\
B^{1 f g} &=& i a (\mh^{f g}-\mh^{g f})
\nonumber\\
B^{f_1 f_2 f_3} &=& i \rd (\bh^{f_1 f_2 f_3} - \bth^{f_1 f_2 f_3})
\nonumber\\
\CB^{1 f_1 f_2 f_3}&=& \rd a (\bh^{f_1 f_2 f_3} + \bth^{f_1 f_2 f_3})
\nonumber\\
\CB^{f_1 f_2 f_3 f_4}&=& -4 \mh^{[f_1 f_2} \mh^{f_3 f_4]}
+4 \rd (s^{[f_1}\bh^{f_2 f_3 f_4]} +s^{[f_1}\bth^{f_2 f_3 f_4]})
\ena
where
\eqa
\mh^{f g}&=& q^{c f} {\bar q}^g_c +\um s^f s^g
\nonumber\\
\bh^{f_1 f_2 f_3}&=& \ee_{c_1 c_2 c_3} q^{c_1 f_1} q^{c_2 f_2} q^{c_3 f_3}
+{3\rd\over 2} s^{[f_1} \mh^{f_2 f_3]}
\nonumber\\
\bth^{f_1 f_2 f_3}&=& \ee_{c_1 c_2 c_3} {\bar q}^{c_1 f_1} {\bar q}^{c_2 f_2}
{\bar q}^{c_3 f_3}
-{3\rd\over 2} s^{[f_1} \mh^{f_2 f_3]}
\ena
where the sign in the last two equations is the right one compatible
with the pseudo conjugation exchanging $q$ with ${\bar q}$.

\end{document}